# A Multi-Source Retrieval Question Answering Framework Based on RAG

First Author: Ridong Wu

Affiliation: School of Computer Science and Network Engineering, Guangzhou University, Guangzhou, China

Email:32106200097@e.gzhu.edu.cn

Third Author: Xiangbiao Su

Affiliation: School of Computer Science and Network Engineering, Guangzhou University, Guangzhou, China

Email:32106500006@e.gzhu.edu.cn

Fifth Author: Yifei Liao

Affiliation: School of Computer Science and Network Engineering, Guangzhou University, Guangzhou, China

Email:32106200102@e.gzhu.edu.cn

Second Author:Shuhong Chen*

Affiliation: School of Computer Science and Network Engineering, Guangzhou University, Guangzhou, China

* Corresponding author:shuhongchen@gzhu.edu.cn

Fourth Author: Yuankai Zhu

Affiliation: School of Computer Science and Network Engineering, Guangzhou University, Guangzhou, China

Email:32106200107@e.gzhu.edu.cn

Sixth Author:Jianming Wu

Affiliation: School of Computer Science and Network Engineering, Guangzhou University, Guangzhou, China

Email:jianmingwu@e.gzhu.edu.cn

*Abstract*—With the rapid development of large-scale language models, Retrieval-Augmented Generation (RAG) has been widely adopted. However, existing RAG paradigms are inevitably influenced by erroneous retrieval information, thereby reducing the reliability and correctness of generated results. Therefore, to improve the relevance of retrieval information, this study proposes a method that replaces traditional retrievers with GPT-3.5, leveraging its vast corpus knowledge to generate retrieval information. We also propose a web retrieval based method to implement fine-grained knowledge retrieval, Utilizing the powerful reasoning capability of GPT-3.5 to realize semantic partitioning of problem.In order to mitigate the illusion of GPT retrieval and reduce noise in Web retrieval,we proposes a multi-source retrieval framework, named MSRAG, which combines GPT retrieval with web retrieval. Experiments on multiple knowledge-intensive QA datasets demonstrate that the proposed framework in this study performs better than existing RAG framework in enhancing the overall efficiency and accuracy of QA systems.

*Keywords*-Large Language Models; Retrieval-Augmented Generation; Web Retrieval; GPT Retrieval; Question-Answering (QA)

## I. INTRODUCTION

Due to their extensive knowledge base, Large Language Models (LLMs) have become indispensable tools for everyday information retrieval[1] . However, LLMs are limited to the knowledge acquired during the pre-training phase and lack real-time updates. To address this challenge, the RAG (Retrieval-Augmented Generation) technique has emerged.

LLMs by integrating information from external knowledge repositories with the generative model's capabilities to improve the accuracy and relevance of question answering.In 2020, Kelvin et al. introduced REALM[2] , which leverages an indexed corpus of paragraphs to support conditional generation. Subsequently, Michael et al. proposed Re2G[3], which integrates neural initial retrieval and re-ranking into RAG, thereby enhancing information relevance. Reiichiro et al. presented WebGPT [4], which fine-tunes GPT-3 and utilizes Web search functionality to acquire knowledge information.

However, traditional Retrieval-Augmented Generation (RAG) has certain limitations. Firstly, the retrieved document passages may not be fully relevant to the query, leading to misinformation or incoherence. Secondly, if relevant passages are not retrieved, the model may fabricate answers due to insufficient context or generate responses that fail to accurately address the query. Overall, relying solely on the initial input from a Large Language Model (LLM) to retrieve relevant information from external corpora is often inadequate for complex multi-step and long-form generation tasks.

To address that problem, Linhao Ye et al.[5] proposed a dialogue-level RAG method, which integrates fine-grained retrieval augmentation and conversational question-answering (CQA) self-checking.Wenhao Yu et al. [6] proposed the substitution of document retrieval systems with large-scale language models for generating retrieval information Weihang Su et al.[7] introduced a novel framework designed to determine when and what to retrieve based on the real-time informational needs of the Language Model (LM) during text generation.In response to advancements in information retrieval techniques, we've optimized the RAG framework by integrating ideas from Wenhao Yu et al. [6] .This includes incorporating a Web retrieval model alongside GPT-based retrieval and non-retrieval strategies. Our approach also integrates a large language model within the Web retrieval module, aiming to enhance information retrieval relevance and reduce noise.

Based on the above analysis, the main contributions of this paper are as follows:

(1) In order to enhance the relevance of retrieved information, this study proposes a method that replaces traditional retrievers with GPT-3.5. Additionally, to improve the granularity of retrieved information, this research proposes a method based on a Web retrieval framework, utilizing the powerful inferential capabilities of GPT-3.5 for semantic segmentation of queries.

(2) In order to mitigate the illusionary impact of GPT retrieval and reduce the noise in Web retrieval, this study proposes a multi-source retrieval approach that combines GPT retrieval with Web retrieval.

(3) In order to evaluate the performance of the multi-source retrieval framework proposed in this study, based on RAG，this study conducted validation experiments on multiple knowledge-intensive QA datasets and found that the proposed multi-source retrieval framework based on RAG outperforms other RAG frameworks in enhancing the overall efficiency and accuracy of QA systems.

## II. SYSTEM MODEL

This paper attenuates information noise in retrieval and enhances the relevance of retrieved information by proposing a multi-source retrieval framework based on RAG.

As illustrated in Figure. 1, in MSRAG, there are GPT-3.5, a Web search engine (Google), and LLM. Initially, within the Web retrieval module, we utilize the Web search engine to perform searches for questions in order to obtain real-time information. Moreover, through experiments, we discovered that simply inputting the original question into the Web search does not yield significantly relevant results(Figure. 2). Therefore, harnessing the robust semantic capabilities of GPT-3.5, we semantically segment the original question into three most relevant and non-repetitive sub-questions, conducting Web searches on a per-sub-question basis. To further mitigate the noise in the retrieved information, we employ GPT-3.5 to summarize the information retrieved from the sub-questions, ultimately deriving the Information-Web.

Secondly, when dealing with complex multi-step questions, Web retrieval often fails to provide satisfactory search results. To address this issue, we replace conventional search engines with GPT-3.5, leveraging its robust semantic understanding capabilities and vast knowledge repository of linguistic contexts to generate search information pertinent to the given queries. Throughout this process, we primarily employ the CoT thinking chain[8] to prompt GPT-3.5 in generating contextual documents based on the provided queries(Figure. 3), thereby acquiring Information-GPT.

Subsequently, due to the potential risk of hallucination associated with GPT-3.5 in generating retrieval information, the framework also directly inputs the original question into LLM. By utilizing the answers generated by all three components, a loss function calculation is performed to select the answer with the lowest loss value as the optimal answer. This approach ensures mutual balance and compensation for each component's respective shortcomings.

The specific details of MSRAG are outlined as follows: (1) Initially, the original question is input into GPT-3.5. Utilizing CoT thought chains[8], the original question undergoes semantic segmentation to generate three most relevant and non-repetitive sub-questions. Contextual document generation is conducted, resulting in Information-GPT. (2) Subsequently, Web retrieval is performed on the sub-questions generated in the previous step to acquire three sets of retrieval information. (3) Utilizing GPT-3.5 once again, the retrieval information obtained in the previous step is merged and summarized to derive Information-Web. (4) Information-Web & original question, Information-GPT & original question, and the original question are then separately input into LLM, yielding answer-Web, answer-GPT, and answer-not. (5) A loss value calculation is conducted on the three answers to determine the final answer. Let's denote "answer-Web," "answer-GPT," and "answer-not" as $A$, $B$, and $C$ respectively, with the correct answer represented as $D$. We will then calculate the cosine similarity for each pair:

$$\text{Cosine similarity}_{A-D} = \frac{A \cdot D}{\|A\| \cdot \|D\|} \quad (1)$$

$$\text{Cosine similarity}_{B-D} = \frac{B \cdot D}{\|B\| \cdot \|D\|} \quad (2)$$

$$\text{Cosine similarity}_{C-D} = \frac{C \cdot D}{\|C\| \cdot \|D\|} \quad (3)$$

To select the highest cosine similarity value as the final score, we compare the calculated cosine similarity values for each pair and choose the maximum. Let's denote the maximum cosine similarity value as Cosine similarity$_{max}$. The expression for this is as follows:

$$\text{Cosine similarity}_{max} = \max\left(\frac{A \cdot D}{\|A\| \cdot \|D\|}, \frac{B \cdot D}{\|B\| \cdot \|D\|}, \frac{C \cdot D}{\|C\| \cdot \|D\|}\right) \quad (4)$$

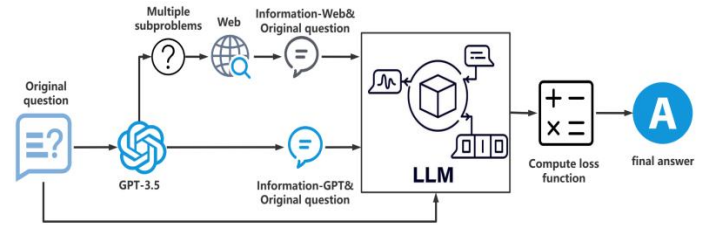

Figure 1. The Framework of MSRAG

Unlike traditional RAG, our proposed MSRAG framework integrates two external knowledge sources, including Web retrieval, GPT retrieval, and non-retrieval. The loss calculation process in the framework involves utilizing a pre-trained BERT model to convert answers into vectors, followed by computing the cosine similarity of each answer. The answer with the highest similarity score is then selected for output. By computing the loss function to obtain the optimal answer, this feature achieves a balanced performance between Web retrieval and GPT retrieval. Additionally, in the Web retrieval module, we introduce the CoT thought chain, fully leveraging the robust semantic capabilities of GPT-3.5 for semantic decomposition, thereby enhancing the granularity of retrieval information. Furthermore, we propose the utilization of GPT-3.5 to replace conventional retrievers, generating the required retrieval information from its extensive corpus knowledge base, thus further improving the effectiveness of retrieval information.

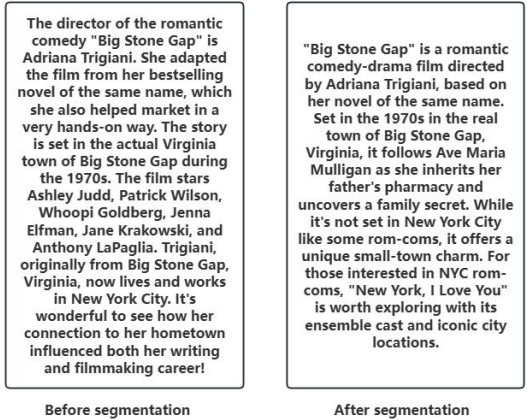

Figure 2. Examples of Before and After Semantic Segmentation Comparison

Figure 3. GPT-Generated Retrieval Prompts

## III. EXPERIMENT

### A. Experiment Setup

In this study, we utilized Python version 3.11.5 as the programming language, GPT-3.5, and Google search engine. The language model employed in this research was LLaMa2-7B-Chat [9]. Furthermore, we opted for two MultiHopQA datasets, 2WikiMultiHopQA[10], and HotpotQA[11] to assess the MSRAG framework's capability in addressing complex questions requiring multi-hop reasoning. Additionally, we employed the StrategyQA[12] dataset to evaluate the commonsense reasoning abilities of MSRAG and other baselines. Considering the randomness of experimental outcomes, we conducted three independent experiments and averaged the results for final analysis.And .In terms of evaluation metrics, we utilize F1 score, exact match (EM), and accuracy (ACC) as assessment standards.

### B. Performance analysis and comparison

To assess the performance of MSRAG, we not only compared it with wo-RAG (which provides answers directly from the LLM) and SR-RAG (which retrieves relevant paragraphs based on the query), but also with several advanced RAG, including FL-RAG[13], which employs previously set markers at intervals as query triggers for retrieval; FS-RAG [14], which utilizes the last generated sentence as a query trigger for retrieval of each sentence; and FLARE [15], which activates retrieval upon encountering uncertain markers, using preceding sentences without uncertainties as queries.

Table 1 presents the performance of each RAG model on the 2WikiMultiHopQA, HotpotQA, and StrategyQA datasets. As shown in the table, MSRAG outperforms other RAGs across all datasets, demonstrating its superior performance in question-answering tasks.

Table 1.0 Performance Comparison of Different Methods on Various Datasets
(**Bolded results** denote the highest performance in each category.)

| Method | 2Wiki | | HotpotQA | | StrategyQA |
|---|---|---|---|---|---|
| | EM | F1 | EM | F1 | Accuracy |
| No-RAG | 0.135 | 0.1282 | 0.108 | 0.1123 | 0.557 |
| SR-RAG | 0.169 | 0.2549 | 0.164 | 0.2499 | 0.654 |
| FL-RAG | 0.112 | 0.1922 | 0.146 | 0.2107 | 0.635 |
| FLARE | 0.143 | 0.2134 | 0.149 | 0.2208 | 0.627 |
| FS-RAG | 0.189 | 0.2652 | 0.214 | 0.3035 | 0.629 |
| MSRAG | **0.508** | **0.5646** | **0.303** | **0.3066** | **0.863** |

To demonstrate the effectiveness of GPT-3.5 as a replacement for conventional retrieval systems, we compared GPT retrieval with No-RAG across the 2WikiMultiHopQA, HotpotQA, and StrategyQA datasets.

Table 2.0 : Performance comparison of GPT retrieval versus No RAG
(**Bolded results** denote the highest performance in each category.)

| Method | 2Wiki | | Hotpot | | StrategyQA |
|---|---|---|---|---|---|
| | EM | F1 | EM | F1 | Accuracy |
| No-RAG | 0.135 | 0.1282 | 0.108 | 0.1123 | 0.557 |
| GPT-Retrieval | **0.172** | **0.1685** | **0.241** | **0.2427** | **0.677** |

Table 2 demonstrates the performance of GPT retrieval versus No RAG on the 2WikiMultiHopQA, HotpotQA, and StrategyQA datasets. As shown in the table, GPT Retrieval outperforms No RAG across all datasets, substantiating the efficacy of the GPT retrieval method.

To demonstrate the effectiveness of our integrated framework, we conducted ablation studies and performed comparative evaluations on the 2WikiMultiHopQA, HotpotQA, and StrategyQA datasets.

**Table3.0** Ablation Experiment
(**Bolded results** denote the highest performance in each category.)

| Method | 2Wiki | | HotpotQA | | StrategyQA |
|---|---|---|---|---|---|
| | EM | F1 | EM | F1 | Accuracy |
| w/o GPT | 0.201 | 0.2013 | 0.186 | 0.1891 | 0.764 |
| w/o Web | 0.254 | 0.2518 | 0.278 | 0.2826 | 0.806 |
| MSRAG | **0.508** | **0.5646** | **0.303** | **0.3066** | **0.863** |

Table 3 presents the performance of MSRAG compared to w/o GPT and w/o Web on the 2WikiMultiHopQA, HotpotQA, and StrategyQA datasets. As shown in the table, MSRAG outperforms both w/o GPT and w/o Web on all datasets, demonstrating the effectiveness of the integrated approach.

In summary, compared to other methods, MSRAG demonstrates superior performance when addressing complex multi-step problems. Additionally, experiments have confirmed the feasibility of using GPT as a substitute for general retrievers, while ablation studies have validated the effectiveness of RAG integration.

## IV. CONCLUSION

This paper proposes a multi-source retrieval framework based on RAG that effectively mitigates the issue of excessive noise in retrieval information by integrating Web retrieval and GPT-Retrieval. It also enhances the granularity and relevance of retrieval information through semantic segmentation of questions. MSRAG significantly outperforms existing RAG methods across various knowledge-intensive benchmark tests. One of the key future research objectives could focus on enhancing the performance of GPT-Retrieval and investigating effective methods to reduce operational costs and accelerate runtime within integrated approaches.

*ACKNOWLEDGMENT*

This work was supported by the Guangzhou University national College Student Innovation Training Program of China under Grant No.202311078034, the Guangdong Provincial Natural Science Foundation under Grant No. 2022A1515011386, and the National Scholarship Fund of China under Grant No. 202308440307 .